\documentclass[aps,prb,preprint,showpacs]{revtex4}
\usepackage{epsfig,amssymb,amsfonts}
\usepackage{graphicx}
\input{epsf}
\begin{document}

\title{A general theory for the van der Waals
interactions in colloidal systems based on fluctuational
electrodynamics}

\author{Vassilios Yannopapas}
\email{vyannop@upatras.gr}
\affiliation{Department of Materials
Science, University of Patras, GR-26504 Patras, Greece}
\date{\today}

\begin{abstract}
A rigourous theory for the determination of the van der Waals
interactions in colloidal systems is presented. The method is
based on fluctuational electrodynamics and a multiple-scattering
method which provides the electromagnetic Green's tensor. In
particular, expressions for the Green's tensor are presented for
arbitrary, finite, collections of colloidal particles, for
infinitely periodic or defected crystals as well as for finite
slabs of crystals. The presented formalism allows for {\it ab
initio} calculations of the vdW interactions is colloidal systems
since it takes fully into account retardation, many-body,
multipolar and near-fields effects.
\end{abstract}

 \maketitle
\bibliographystyle{apsrev}

\section{Introduction}
\label{intro}

The van der Waals (vdW) interactions are particularly important in
colloidal systems since, along with the electrostatic forces, they
determine the structure of such systems. The stability of
colloidal systems resulting from the interplay between the vdW and
the electrostatic interactions is very well elucidated in the
context of Derjaguin-Landau-Verwey-Overbeek theory. \cite{dlvo}
The vdW interactions which originate from the irreducible
electromagnetic (EM) fluctuations of vacuum are usually calculated
by means of the Hamaker approach, \cite{hamaker} where the force
stems from simple pairwise addition of the corresponding
intermolecular forces, \cite{israelachvili,parsegian_book}
although the vdW interactions are not additive. A rigorous
treatment of the vdW interactions based on fluctuational
electrodynamics \cite{rytov,agarwal} has been pioneered by
Lifshitz \cite{lifshitz} for the case of two infinite half-spaces.
The Lifshitz theory has been extended to the case of pairs of
finite-sized objects such as spheres or cylinders (Derjaguin
approximation) \cite{parsegian_book,rajter} which is valid,
however, for very short distances between the objects, in the
nonretarded limit. In some cases, elements of the Lifshitz theory
for half-spaces are incorporated within the Hamaker formula for
the vdW force between two particles, in the form of semi-empirical
corrections.
\cite{gregory,pailthorpe,israelachvili,parsegian_book,prieve} By
use of perturbation theory and the Clausius - Mossotti formula,
Langbein \cite{langbein,langbein_book} developed a general
formalism for the vdW force between two spheres which has been
primarily applied to aerosol particles.
\cite{marlow_emp,arunachalam,marlow_ret}

Recently, a new, rigorous theory based on fluctuational
electrodynamics for the calculation of the vdW interactions among
a collection of macroscopic bodies of finite size has been
proposed. \cite{yv_vdw} This theory is based on a
multiple-scattering Green's tensor formalism incorporated within
the framework of fluctuational electrodynamics. More specifically,
the vdW force results from the integration over the surface of the
bodies of the Maxwell stress tensor of the vacuum/ thermal EM
field which is provided by the fluctuation-dissipation theorem and
through this by the Green's tensor of the classical EM field. The
calculation of the Green's tensor is based on an EM multiple
scattering formalism for arbitrary collections of scatterers. The
multiple-scattering Green's tensor formalism offers a precise
knowledge of the fluctuating EM field by going beyond the
approximation of pairwise interactions between the scatterers and
by taking into account the full multipole interactions between
them. Furthermore, since it constitutes a solution to the
inhomogeneous wave equation, retardation effects are included {\em
a priori} in the presented formalism. In addition, metallic and
dielectric particles are treated on an equal footing since the
method in question also accounts for the magnetic-field vacuum
fluctuations which cannot be neglected in the case of metallic
particles. Finally, the effect of finite temperature can be easily
addressed. We note that a different approach has been recently
presented \cite{rodriguez} where the EM Green's tensor entering
the fluctuation-dissipation theorem is calculated by means of a
finite-difference frequency-domain method.

When a particle is a member of a colloidal crystal and a net vdW
force exerted on the particle is evident (e.g., in a finite slab
of a colloidal crystal or in an infinite crystal containing point
and/ or line defects), it is calculated from a pairwise addition
of the forces stemming from the all the other particles of the
crystal. So, at first glance, an extension of the
Ref.~\onlinecite{yv_vdw} to the case of a colloidal system would
be based on a pairwise summation of the (exact) force for a pair
of particles. However, such an approach is only approximately
correct since the vdW interactions are not additive. The way to
extend the method of Ref.\onlinecite{yv_vdw} to the case of a
colloidal crystal is to derive a semi-analytical expression of the
EM Green's tensor for the particular crystal. The knowledge of the
EM Green's tensor everywhere in space allows the calculation of
the cross-spectral correlation functions of the vacuum EM field
which are contained in the EM Maxwell stress tensor, by
application of the fluctuation-dissipation theorem. By integrating
the Maxwell stress tensor over the surface of the particle we
obtain the vdW force. The paper is organized as follows. In
section \ref{sec:fd} we provide a brief overview of fluctuational
electrodynamics and the Maxwell stress tensor. In section
\ref{sec:green_tensor} we provide expressions for the EM Green's
tensor, (a) for arbitrary collections of a finite number of
scatterers, (b) for infinite, periodic and defected crystals, and
(c) for finite slabs of colloidal crystals. In section
\ref{sec:example} we apply the formalism to the case of a
monolayer of polystyrene spheres containing a single defect.
Section \ref{sec:conclusion} concludes the paper.

\section{van der Waals force}
\label{sec:fd}

\subsection{Maxwell stress tensor}
\label{mst}

We consider a finite scatterer with electric permittivity
$\epsilon_{s}$ and/or magnetic permeability $\mu_{s}$ different
from those, $\epsilon_h,\ \mu_h$ of the surrounding homogeneous
medium. According to classical electrodynamics, the exerted force
${\bf F}$ on a finite scatterer in the presence of electric ${\bf
E}$ and magnetic ${\bf H}$ fields satisfying the Maxwell equations
is obtained by integrating the time-average Maxwell stress tensor
${\rm T}_{ij}$ \cite{jackson} over the surface around the
scatterer
\begin{equation}
\langle F_{i} \rangle_{t}= \int_{S} \sum_{j} \langle {\rm T}_{ij}
\rangle_{t} n_{j} dS  \label{eq:force}
\end{equation}
where $\langle .. \rangle_{t}$ denotes the time average, ${\bf n}$
is the normal vector at the surface surrounding the object, and
$i,j=x,y,z$. The components of the tensor $\langle {\rm T}_{ij}
\rangle_{t}$ are given by
\begin{eqnarray}
\langle {\rm T}_{ij} \rangle_{t}&=& \epsilon_{h} \epsilon_{0}
\langle E_{i}({\bf r},t)E_{j}({\bf r},t) \rangle_{t} +\mu_{h}
\mu_{0}\langle H_{i}({\bf r},t)H_{j}({\bf r},t) \rangle_{t}
\nonumber \\
&-&\frac{1}{2} \delta_{ij} \bigl[\epsilon_{h} \epsilon_{0}
\sum_{i'} \langle E_{i'}({\bf r},t)E_{i'}({\bf r},t) \rangle_{t}+
\mu_{h}\mu_{0} \sum_{i'} \langle H_{i'}({\bf r},t)H_{i'}({\bf
r},t) \rangle_{t} \bigr]. \label{eq:mst}
\end{eqnarray}
$\delta_{ij}$ is the Kronecker symbol and $\epsilon_{0}, \mu_{0}$
are the electric permittivity and magnetic permittivity of vacuum,
respectively.

\subsection{Fluctuation-dissipation theorem}
\label{fluct}

In the absence of other radiation sources, the fields ${\bf E},
{\bf H}$ are generated by the thermal radiation emitted from the
same or neighboring scatterers at finite temperature (thermal
fluctuations) or by vacuum radiation at zero temperature
(zero-point fluctuations). The time-correlation function $\langle
E_{i}({\bf r},t+\tau) E_{j}({\bf r}',t) \rangle_{t}$ contained in
Eq.~(\ref{eq:mst}) is calculated within the framework of
fluctuational electrodynamics, \cite{agarwal,rytov} namely from
\cite{joulain}
\begin{equation}
\langle E_{i}({\bf r},t+\tau)E_{j}({\bf r}',t) \rangle_{t}={\rm
Re} \Biggl[ \int_{0}^{\infty} \frac{d \omega}{2 \pi} \exp({\rm i}
\omega \tau) W_{ij}^{EE}({\bf r},{\bf r}';\omega) \Biggr].
\label{eq:time_corr_E}
\end{equation}
The quantity $W_{ij}^{EE}({\bf r},{\bf r}';\omega)$ is the
cross-spectral correlation function for the electric field. For a
system at thermal equilibrium, i.e., the scatterer, the
surrounding medium and its neighbouring scatterers at the same
temperature $T$, $W_{ij}$ is provided by the
fluctuation-dissipation theorem \cite{agarwal,rytov,joulain}
\begin{equation}
W_{ij}^{EE}({\bf r},{\bf r}';\omega)=4 \omega \mu_{h} \mu_{0} c^2
{\rm Im} G_{ij}^{EE} ({\bf r},{\bf r}';\omega) \hbar \omega
\Bigl[1+\frac{1}{\exp (\hbar \omega/k_{B} T)-1}\Bigr],
\label{eq:wee}
\end{equation}
where $\hbar$ is the reduced Planck's constant, $k_{B}$ is the
Boltzmann's constant and $G^{EE}_{ij}({\bf r},{\bf r}';\omega)$ is
the component of the full Green's tensor $G_{ij}$ which provides
the electric field at ${\bf r}$ due to an electric dipole source
at ${\bf r}'$. The time-correlation function $\langle H_{i}({\bf
r},t+\tau) H_{j}({\bf r}',t) \rangle_{t}$ for the magnetic field
is given similar to Eq.~(\ref{eq:time_corr_E}) with $W_{ij}^{EE}$
substituted by
\begin{equation}
W_{ij}^{HH}({\bf r},{\bf r}';\omega)=4 \omega \epsilon_{h}
\epsilon_{0} c^2 {\rm Im} G_{ij}^{HH} ({\bf r},{\bf r}';\omega)
\hbar \omega \Bigl[1+\frac{1}{\exp (\hbar \omega/k_{B}
T)-1}\Bigr]. \label{eq:whh}
\end{equation}
We note that, the final value of the vdW force acting on a
scatterer is obtained by subtracting from Eq.~(\ref{eq:force}) the
force which remains in the absence of the scatterer as it is the
case for the calculation of the Casimir force between two
semi-infinite slabs. \cite{schwinger} However, in vacuum, the
Green's tensor and the corresponding Maxwell stress tensor,
Eq.~(\ref{eq:mst}), are constant in space and their integral over
a closed surface is zero. From the above, it is obvious that the
central quantity which essentially determines the force acting on
the scatterer is the EM Green's tensor.

\section{Electromagnetic Green's tensor}
\label{sec:green_tensor}

\subsection{Multipole expansion of the EM field}
\label{subsec:multi}

Let us consider a harmonic EM wave, of angular frequency $\omega$
which is described by its electric-field component
\begin{equation}
{\bf E}({\bf r},t)={\rm Re}\left[{\bf E}({\bf r}){\rm exp}(-{\rm
i}\omega t)\right]\,. \label{eq:harmonicelectric}
\end{equation}
In a homogeneous medium characterized by a dielectric function
$\epsilon(\omega)\epsilon_0$ and a magnetic permeability
$\mu(\omega)\mu_0$, where $\epsilon_0$, $\mu_0$ are the electric
permittivity and magnetic permeability of vacuum, Maxwell
equations imply that ${\bf E}({\bf r})$ satisfies a vector
Helmholtz equation, subject to the condition $\nabla \cdot {\bf
E}=0$, with a wave number $q=\omega/c$, where $c=1/\sqrt{\mu
\epsilon \mu_0 \epsilon_0}=c_{0}/\sqrt{\mu \epsilon}$ is the
velocity of light in the medium. The spherical-wave expansion of
${\bf E}({\bf r})$ is given by \cite{jackson}
\begin{equation}
{\bf E}({\bf r})=\sum_{l=1}^{\infty}\sum_{m=-l}^{l}
 \left\{a_{E l m}
f_{l}(qr){\bf X} _{l m}({\bf \hat r})+a_{E l m}\frac{{\rm
i}}{q}\nabla\times \left[f_{l}(qr){\bf X} _{l m}({\bf \hat
r})\right]\right\}\;, \label{eq:multem}
\end{equation}
where $a^P_{l m}$ ($P=E,H$) are coefficients to be determined.
${\bf X} _{l m}({\bf \hat r})$ are the so-called vector spherical
harmonics \cite{jackson} and $f_{l}$ may be any linear combination
of the spherical Bessel function, $j_{l}$, and the spherical
Hankel function, $h_{l}^{+}$. The corresponding magnetic
induction, ${\bf B}({\bf r})$, can be readily obtained from ${\bf
E}({\bf r},t)$ using Maxwell's equations \cite{jackson}.

\subsection{Scattering from a single scatterer and the corresponding Green's tensor}
\label{subsec:single}

In this subsection we present a brief summary of the solution to
the problem of EM scattering from a single sphere (Mie scattering
theory \cite{jackson,bohren}) along with the expression for the
single-sphere Green's tensor. We consider a sphere of radius $S$,
with its center at the origin of coordinates, and assume that its
electric permittivity $\epsilon_{s}$ and/or magnetic permeability
$\mu_{s}$ are different from those, $\epsilon_h,\ \mu_h$ of the
surrounding homogeneous medium. An EM plane wave incident on this
scatterer is described, respectively, by Eq.~(\ref{eq:multem})
with $f_{l}=j_{l}$ (since the plane wave is finite everywhere) and
appropriate coefficients $a_L^0$, where $L$ denotes collectively
the indices $P l m$. That is,
\begin{equation}
{\bf E}^{0}({\bf r})=\sum_{L} a_{L}^{0} {\bf J}_{L}({\bf r})
\label{eq:sph_inc}
\end{equation}
where
\begin{equation}
{\bf J}_{E l m}({\bf r})=\frac{{\rm i}}{q_h} \nabla \times
j_{l}(q_h r) {\bf X}_{l m}(\hat{{\bf r}}), \ \ \ {\bf J}_{H l
m}({\bf r})=j_{l}(q_h r) {\bf X}_{l m}(\hat{{\bf r}})
\label{eq:j_func}
\end{equation}
and $q_h=\sqrt{\epsilon_h \mu_h} \omega /c_{0}$. The coefficients
$a_L^0$ depend on the amplitude, polarization and propagation
direction of the incident EM plane wave and are given by
Eqs.~(\ref{sphinp}) (subsection \ref{pcalc}) for ${\bf g}={\bf
0}$.

Similarly, the wave that is scattered from the sphere is described
by Eq.~(\ref{eq:multem}) with $f_{l}=h^+_{l}$, which has the
asymptotic form appropriate to an outgoing spherical wave:
$h^+_{l}\approx (-{\rm i})^{l}\exp({\rm i} q_h r)/{\rm i} q_h r$
as $r\rightarrow\infty$, and appropriate expansion coefficients
$a^+_L$. Namely,
\begin{equation}
{\bf E}^{+}({\bf r})=\sum_{L} a_{L}^{+} {\bf H}_{L}({\bf r})
\label{eq:sph_sc}
\end{equation}
where
\begin{equation}
{\bf H}_{E l m}({\bf r})=\frac{{\rm i}}{q_h} \nabla \times
h^{+}_{l}(q_h r) {\bf X}_{l m}(\hat{{\bf r}}), \ \ \ {\bf H}_{H l
m}({\bf r})=h^{+}_{l}(q_h r) {\bf X}_{l m}(\hat{{\bf r}}).
\label{eq:h_func}
\end{equation}

The wavefield for $r>S$ is the sum of the incident and scattered
waves, i.e., ${\bf E}^{out}={\bf E}^{0}+{\bf E}^{+}$. By applying
the requirement that the tangential components of ${\bf E}$ and
${\bf H}$ be continuous at the surface of the scatterer, we obtain
a relation between the expansion coefficients of the incident and
the scattered field, as follows:
\begin{equation}
a^+_L=\sum_{L'} T_{LL'}\, a^0_{L'}\;, \label{eq:tmatrix}
\end{equation}
where $T_{LL'}$ are the elements of the so-called scattering
transition $T$-matrix. \cite{bohren} Eq.~(\ref{eq:tmatrix}) is
valid for any shape of scatterer; explicit relations of the
$T$-matrix for scatterers of various shapes can be found
elsewhere. \cite{doicu,nonsph}

The Green's tensor for a single sphere is given by \cite{yv_ldos}
\begin{equation}
G^{(s)}_{ii'}({\bf r},{\bf r}')=-{\rm i}\omega \frac{(\epsilon_h
\mu_h)^{3/2}}{c_{0}^{3}} \sum_{L} [R_{L;i}({\bf r})
\overline{I}_{L;i'} ({\bf r}') \Theta(r'-r) + I_{L;i}({\bf r})
\overline{R}_{L;i'} ({\bf r}') \Theta(r-r')]
\label{eq:green_sphere}
\end{equation}
The vector functions $R_{L;i}({\bf r}), \overline{R}_{L;i}({\bf
r})$ are dimensionless eigenfunctions of the wave operator
\begin{equation}
{\bf \Lambda}({\bf r})=\frac{c_0^{2}}{\epsilon({\bf r}) \mu({\bf
r})} \nabla \times \nabla \times  \label{eq:lambda_op}
\end{equation}
for a single scatterer which are regular at its center.
\cite{yv_ldos,sainidou_green} The vector functions $I_{L;i}({\bf
r}), \overline{I}_{L;i}({\bf r})$ are also eigenfunctions of the
operator (\ref{eq:lambda_op}) but they are infinite at the sphere
center. \cite{yv_ldos,sainidou_green} The Green's tensor of
Eq.~(\ref{eq:green_sphere}) will be the basis for the construction
of the corresponding tensor for a collection of spheres.

\subsection{Green's tensor for many scatterers}
\label{subsec:many}

We consider a collection of $N$ nonoverlapping scatterers
described by a permittivity $\epsilon_{s}$ and permeability
$\mu_{s}$ centred at sites ${\bf R}_{n}$ in a homogeneous host
medium described by $\epsilon_h$, $\mu_h$, respectively. In
site-centered representation, the Green's tensor for the system of
scatterers satisfies \cite{yv_ldos,sainidou_green}
\begin{equation}
\sum_{i}\left[\omega^{2} \delta_{i''i} - \Lambda_{i''i}({\bf
R}_{n}+{\bf r}_{n}) \right] G_{ii'}({\bf R}_{n}+{\bf r}_{n},{\bf
R}_{n'}+{\bf r'}_{n'})=\delta_{i''i'} \delta({\bf r}_{n}-{\bf
r'}_{n'}) \delta_{nn'} \label{eq:wave_source_site}
\end{equation}
where ${\bf r}_{n}={\bf r}-{\bf R}_{n}$, ${\bf r'}_{n'}={\bf
r'}-{\bf R}_{n'}$, and $i,i'=x,y,z$. The operator
$\Lambda_{i''i}({\bf r})$ is given by Eq.~(\ref{eq:lambda_op}). It
can be verified that the Green's tensor satisfying
Eq.~(\ref{eq:wave_source_site}) is the following
\cite{yv_ldos,sainidou_green}
\begin{equation}
G_{ii'}({\bf R}_{n}+{\bf r}_{n},{\bf R}_{n'}+{\bf r'}_{n'})=
G^{(s)n}_{ii'}({\bf r}_{n},{\bf r'}_{n'}) \delta_{nn'} -{\rm
i}\omega \frac{(\epsilon_h \mu_h)^{3/2}}{c^{3}} \sum_{LL'}
\overline{R}^{n}_{L;i}({\bf r}_{n}) D^{n'n}_{L'L}
R^{n'}_{L';i'}({\bf r'}_{n'}). \label{eq:green_coll}
\end{equation}
$G^{(s)n}_{ii'}({\bf r}_{n},{\bf r'}_{n'})$ is the Green's tensor
for a single scatterer located at ${\bf R}_{n}$ and it is given by
Eq.~(\ref{eq:green_sphere}). The vector functions
$R^{n}_{L;i}({\bf r}_{n}), \overline{R}^{n}_{L;i}({\bf r}_{n})$
are the dimensionless eigenfunctions of the operator of
Eq.~(\ref{eq:lambda_op}) for the sphere at ${\bf R}_{n}$.
$D_{LL'}^{nn'}$ are propagator functions that represent the
contributions of all possible paths by which a wave outgoing from
the $n'$-th scatterer produces an incident wave on the $n$-th
scatterer, after scattering in all possible ways (sequences) by
the scatterers at all sites including the $n$-th and $n'$-th
scatterers. The specific form of the $D_{LL'}^{nn'}$ propagator
functions depends on the geometrical arrangement of the
scatterers.

\subsection{Propagator for an arbitrary collection of scatterers}
\label{arb}

For an arbitrary collection of a finite number $N$ of scatteres,
the $D$-propagator is given by \cite{yv_ldos}
\begin{equation}
D^{nn'}_{LL'}=\Omega^{nn'}_{LL'}+\sum_{n''} \sum_{L''} \sum_{L'''}
D^{nn''}_{LL''} T^{n''}_{L''L'''} \Omega^{n''n'}_{L'''L'}.
\label{eq:d_propagator}
\end{equation}
The matrix $\Omega^{nn'}_{LL'}$ appearing in
Eq.~(\ref{eq:d_propagator}) is called free-space propagator and
transforms an outgoing vector spherical wave about ${\bf R}_{n'}$
in a series of incoming vector spherical waves around ${\bf
R}_{n}$. \cite{yv_ldos}  The matrix $T^{n}_{LL'}$ is the
scattering $T$-matrix of a scatterer of general shape,
\cite{doicu,nonsph} located at ${\bf R}_{n}$.

\subsection{Propagator for periodic arrays of scatterers}
\label{periodic}

For the case of an infinite number of same spheres arranged
periodically, in one- (1D), two- (2D) or three (3D) dimensions,
the propagator $D^{nn'}_{LL'}$ is given as a Fourier transform
\begin{equation}
D^{nn'}_{LL'}=\frac{1}{v} \int_{BZ} d^{q} k \exp[{\rm i} {\bf k}
\cdot ({\bf R}_{n}-{\bf R}_{n'})] D_{LL'}({\bf k}),
\label{eq:d_propagator_fourier}
\end{equation}
where $q$ is the space dimensionality, the integration in
Eq.~(\ref{eq:d_propagator_fourier}) is carried out within the
Brillouin Zone (BZ), ${\bf k}$ is the Bloch wavevector, and $v$ is
the BZ volume. ${\bf R}_{n}$ are the Bravais lattice vectors.
$D_{LL'}({\bf k})$ is given by
\begin{equation}
D_{LL'}({\bf k})=\Omega_{LL'}({\bf k}) + \sum_{L''L'''}
D_{LL''}({\bf k}) T_{L''L'''} \Omega_{L'''L'}({\bf k}).
\label{eq:d_propagator_dyson}
\end{equation}
$T_{L''L'''}$ is the $T$-matrix of the spheres. $\Omega_{LL'}({\bf
k})$ depend only on the crystal lattice and are known as structure
constants a term which is common in the Korringa-Kohn-Rostoker
method \cite{kkr} for the calculation of the electronic band
structure of atomic solids. They can be found by Ewald-summation
techniques. \cite{ham,moroz_jpa} Eqs.~(\ref{eq:wee}) and
(\ref{eq:whh}) require the calculation of the Green's tensor [via
Eq.~(\ref{eq:green_coll})] for an infinitely periodic lattice of
scatterers; therefore, only the $D^{00}_{LL'}$ component (that for
the central unit cell) is needed since all spheres are equivalent
for the case of a Bravais lattice with one sphere per unit cell.

We note that the propagator of Eq.~(\ref{eq:d_propagator_dyson})
does not yield a net nonzero vdW force, since it corresponds to an
infinitely periodic system. However, the propagator of
Eq.~(\ref{eq:d_propagator_dyson}) can be used as a basis for
calculating the corresponding propagator of a system containing,
e.g., one or more point defects (not symmetrically distributed
within the crystal), in which case a net vdW force emerges. If,
for example, the colloidal particles (described by a a scattering
matrix $T^{n}_{0LL'}$) positioned at ${\bf R}_{n}$ in an otherwise
periodic crystal, are substituted by other, different particles,
each of them described by a scattering matrix $T^{n}_{LL'}$, the
propagator of the defected system is given similar to
Eq.~(\ref{eq:d_propagator}), i.e.,
\begin{equation}
D^{nn'}_{LL'}=D^{nn'}_{0LL'}+\sum_{n''} \sum_{L''} \sum_{L'''}
D^{nn''}_{LL''} \Delta T^{n''}_{L''L'''} D^{n''n'}_{0L'''L'}.
\label{eq:d_propagator_defect}
\end{equation}
where $\Delta T^{n''}_{L''L'''}=T^{n''}_{L''L'''} - T^{n}_{0LL'}$
and $D^{nn'}_{0LL'}$ is the propagator of the periodic system
given by Eqs.~(\ref{eq:d_propagator_fourier}) and
(\ref{eq:d_propagator_dyson}).

\subsection{Propagator for finite slabs}
\label{slab}

In reality, the colloidal systems are not infinitely periodic but
they are actually slabs consisting of a finite number of planes of
particles (scatterers). In this case, the vdW force exerted on a
given scatterer depends on the position of the plane within which
it is located and can therefore be very different for a scatterer
on a surface plane than a scatterer at an innermost plane. In the
following lines, we will provide a formalism for the propagator
for a slab consisting of $N_{p}$ planes of scatterers. It is
assumed that all the planes of the slab have the same 2D
periodicity with the associated lattice vectors given by
\begin{equation}
{\bf R}_{n}=n_{1}{\bf a}_{1}+n_{2}{\bf a}_{2},
 \label{2ddir}
\end{equation}
where ${\bf a}_{1}$ and ${\bf a}_{2}$ are primitive vectors in the
$x y$ plane and $n_{1},n_{2}=0,\pm1,\pm2,\pm3,\cdots$. The
corresponding 2D reciprocal lattice is defined by
\begin{equation}
{\bf g}=m_{1}{\bf b}_{1}+m_{2}{\bf b}_{2}
 \label{2drec}
\end{equation}
where $m_{1},m_{2}=0,\pm1,\pm2,\pm3,\cdots$ and ${\bf b}_{1},{\bf
b}_{2}$ are primitive vectors defined by
\begin{equation}
{\bf b}_{i}\cdot{\bf a}_{j}=2\pi\delta_{ij}\ , \ i,j=1,2.
\end{equation}
Although each plane of the slab must have the same 2D periodicity,
the spheres within each of the $N_{p}$ planes can be different in
terms of shape, size or refractive index.

The propagator for a scatterer residing at the $\nu$-th plane
($\nu=1,2,\cdots,N_{p}$) of a slab is written as a sum of three
terms \cite{cpa3d}

\begin{equation}
F^{00}_{\nu;LL'}=D^{00}_{\nu;LL'} + \sum_{n}\sum_{L''} \sum_{L'''}
P^{0n}_{\nu;LL''}T_{\nu;L''L'''}D^{n0}_{\nu;L'''L'}+P^{00}_{\nu;LL'}
\label{gdef}
\end{equation}

The matrix $D^{n m}_{\nu;LL'}$ represents all the possible
scattering paths {\it within} the $\nu$-th plane by which a wave
outgoing from the $m$-th sphere of this plane produces an incident
wave on the $n$-th sphere of the same plane, after scattering in
all possible ways by all the spheres of this plane including the
central sphere (every sphere represented by the scattering matrix
$T_{\nu:LL'}$). It is given by application of
Eq.~(\ref{eq:d_propagator_fourier}) to a 2D periodic lattice,
i.e.,

\begin{equation}
D^{nm}_{\nu;LL'}=\frac{1}{S_{0}}\int\int_{SBZ}d^{2}k_{\parallel}
\exp({\rm i}{\bf k}_{\parallel}\cdot{\bf R}_{nm}) D_{\nu;LL'}({\bf
k}_{\parallel}) \label{dnm}
\end{equation}
where
\begin{equation}
D_{\nu;LL'}({\bf k}_{\parallel})=\sum_{L''}\Bigr[[{\bf I}-{\bf
\Omega}({\bf k}_{\parallel}){\bf T}_{\nu}]^{-1}\Bigl]_{LL''}
\Omega_{L''L'}({\bf k}_{\parallel}) \label{dfourier}
\end{equation}
where ${\bf R}_{n m}={\bf R}_{n}-{\bf R}_{m}$, $S_{0}$ is the area
of the Surface Brillouin Zone (SBZ) corresponding to
Eq.~(\ref{2drec}), and $\Omega_{LL'}({\bf k}_{\parallel})$ are the
2D structure constants.

The matrix $P^{0n}_{\nu;LL'}$ appearing in the second and third
terms of Eq.~(\ref{gdef}) represents all scattering paths by which
an outgoing wave from the $n$-th sphere of the $\nu$-th plane {\it
exits} from that plane to produce an incident wave on the central
sphere of the same plane after scattering in all possible ways by
all the planes of spheres of the slab, including the $\nu$-th
plane. In the next subsection we will present a summary of the
derivation of $P^{0n}_{\nu;LL'}$ and $ F^{00}_{\nu;LL'}$ which is
given in detail in Ref.~\onlinecite{cpa3d}.

\subsection{Calculation of $P^{0n}_{\nu;LL'}$ and $
F^{00}_{\nu;LL'}$}
 \label{pcalc}

\begin{figure}[h]
\centerline{\includegraphics[width=12cm]{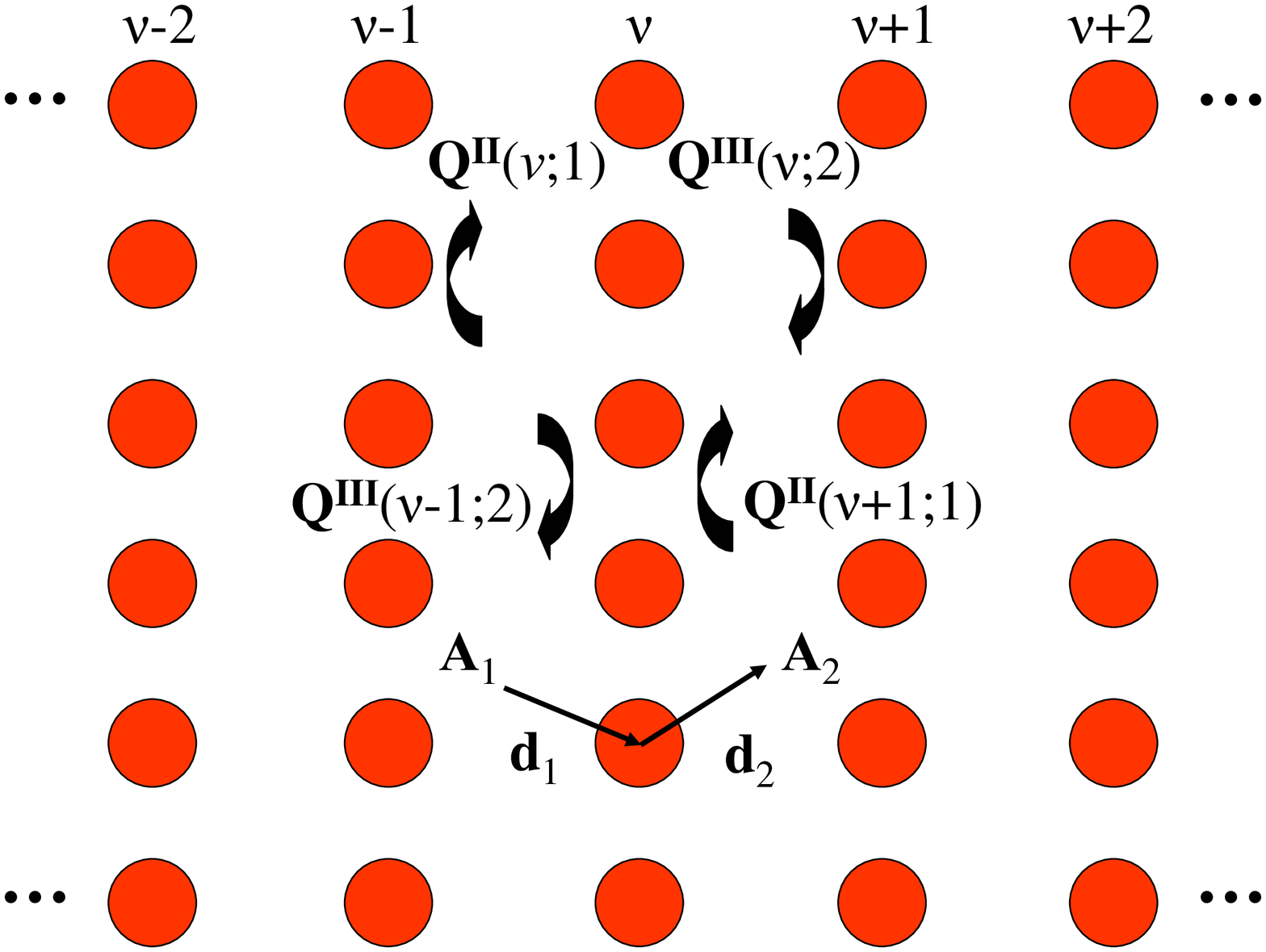}} \caption{The
${\bf Q}$-matrices appearing in Eq.~(\ref{glm}). The position
vectors ${\bf d}_{1},\ {\bf d}_{2}$ of the $\nu$-th layer along
with the corresponding origins ${\bf A}_{1},\ {\bf A}_{2}$ are
also shown.} \label{fig1}
\end{figure}

A wave outgoing from the $n$-th sphere of the $\nu$-th plane has
the form of Eq.~(\ref{eq:sph_sc})
\begin{equation}
{\bf E}^{sc}({\bf r})= \sum_{L} b^{+}_{L}(n;\nu) {\bf H}_{L}({\bf
r}) \label{sc}
\end{equation}
where ${\bf r}_{n \nu}$ is the position vector with respect to the
center of the $n$-th sphere of the $\nu$-th plane. We can expand
the wave of Eq.~(\ref{sc}) into a sum of plane waves propagating
or decaying away from the $\nu$-th plane as follows. \cite{cpa3d}
To the right of the $\nu$-th plane we have
\begin{equation}
{\bf E}^{out\ +}({\bf
r})=\frac{1}{S_{0}}\int\int_{SBZ}d^{2}k_{\parallel}\sum_{{\bf
g}}{\bf E}^{out\ +}_{{\bf g}}({\bf k}_{\parallel}) \exp[{\rm
i}{\bf K}^{+}_{{\bf g}}\cdot({\bf r}-{\bf A}_{2}(\nu))]
\label{outp}
\end{equation}
with
\begin{equation}
E^{out\ +}_{{\bf g};i}({\bf k}_{\parallel})= \exp[-{\rm i}({\bf
k}_{\parallel}\cdot{\bf R}_{n}-{\bf K}^{+}_{{\bf g}}\cdot{\bf
d}_{2}(\nu))]\sum_{L} \Delta_{L;i}({\bf K}^{+}_{{\bf g}})
b^{+}_{L}(n;\nu) \label{ampoutp}
\end{equation}
where $i=1,2$. ${\bf A}_{2}(\nu)$ is a reference point on the
right of the $\nu$-th plane at ${\bf d}_{2}(\nu)$ from its center
(see Fig.~\ref{fig1}). To the left of the $\nu$-th plane we have
\begin{equation}
{\bf E}^{out\ -}({\bf
r})=\frac{1}{S_{0}}\int\int_{SBZ}d^{2}k_{\parallel}\sum_{{\bf
g}}{\bf E}^{out\ -}_{{\bf g}}({\bf k}_{\parallel}) \exp[{\rm
i}{\bf K}^{-}_{{\bf g}}\cdot({\bf r}-{\bf A}_{1}(\nu))]
\label{outm}
\end{equation}
with
\begin{equation}
E^{out\ -}_{{\bf g};i}({\bf k}_{\parallel})= \exp[-{\rm i}({\bf
k}_{\parallel}\cdot{\bf R}_{n}+{\bf K}^{-}_{{\bf g}}\cdot{\bf
d}_{1}(\nu))]\sum_{L} \Delta_{L;i}({\bf K}^{-}_{{\bf g}})
b^{+}_{L}(n;\nu) \label{ampoutm}
\end{equation}
where ${\bf A}_{1}(\nu)$ is a reference point to the left of the
$\nu$-th plane at $-{\bf d}_{1}(\nu)$ from its center (see
Fig.~\ref{fig1}). $\mathbf{K}_{\mathbf{g}}^{\pm}$ is given by
$\mathbf{K}_{\mathbf{g}}^{\pm}= \left(
\mathbf{k}_{\parallel}+\mathbf{g},\ \pm \left[
q^{2}-\left(\mathbf{k}_{\parallel}+\mathbf{g}\right)^{2} \right]
^{1/2} \right)$, where the $+,-$ sign defines the sign of the $z$
component of the wavevector. The coefficients $\Delta_{L;i}$ are
given from Eqs.~(19) and (20) of Ref.~\onlinecite{comphy}.

The plane waves of Eq.~(\ref{outp}) will be multiply reflected
between two parts of the slab, the first (right part) consisting
of all planes to the right of the $\nu$-th plane, and the second
(left part) consisting of all planes to the left of the ($\nu +
1$)-th plane (including the $\nu$-th plane), to produce a set of
plane waves incident on the $\nu$-th plane from the right, which
we can write formally as follows
\begin{equation}
{\bf E}^{in\ -}({\bf
r})=\frac{1}{S_{0}}\int\int_{SBZ}d^{2}k_{\parallel}\sum_{{\bf
g}}{\bf E}^{in\ -}_{{\bf g}}({\bf k}_{\parallel}) \exp[{\rm i}{\bf
K}^{-}_{{\bf g}}\cdot({\bf r}-{\bf A}_{2}(\nu))] \label{inp}
\end{equation}
with
\begin{equation}
E^{in\ -}_{{\bf g};i}({\bf k}_{\parallel})=\sum_{{\bf
g}^{\prime},i^{\prime}} \left\{{\bf Q}^{{\rm III}}(\nu;2)[{\bf
I}-{\bf Q}^{{\rm II}}(\nu+1;1){\bf Q}^{{\rm
III}}(\nu;2)]^{-1}\right\}_{{\bf g}i;{\bf
g}^{\prime}i^{\prime}}E^{out\ +}_{{\bf
g}^{\prime};i^{\prime}}({\bf k}_{\parallel}) \label{ampinp}
\end{equation}
where ${\bf Q}^{{\rm II}}(\nu+1;1)$ and ${\bf Q}^{{\rm
III}}(\nu;2)$ are the appropriate matrices which determine the
reflection (diffraction) of a plane wave by the left and the right
parts of the slab respectively, as defined above. These matrices
are shown schematically in Fig.~\ref{fig1}.

Similarly, the plane waves of Eq.~(\ref{outm}) will be multiply
reflected between two parts of the slab, the first (left part)
consisting of all planes to the left of the $\nu$-th plane and the
second (right part) consisting of all planes to the right of the
($\nu -1$)-th plane (including the $\nu$-th plane), to produce a
set of plane waves incident on the $\nu$-th plane from the left,
which we can write formally as follows
\begin{equation}
{\bf E}^{in\ +}({\bf
r})=\frac{1}{S_{0}}\int\int_{SBZ}d^{2}k_{\parallel}\sum_{{\bf
g}}{\bf E}^{in\ +}_{{\bf g}}({\bf k}_{\parallel}) \exp[{\rm i}{\bf
K}^{+}_{{\bf g}}\cdot({\bf r}-{\bf A}_{1}(\nu))] \label{inm}
\end{equation}
with
\begin{equation}
E^{in\ +}_{{\bf g};i}({\bf k}_{\parallel})=\sum_{{\bf
g}^{\prime},i^{\prime}} \left\{{\bf Q}^{{\rm II}}(\nu;1)[{\bf
I}-{\bf Q}^{{\rm III}}(\nu-1;2){\bf Q}^{{\rm
II}}(\nu;1)]^{-1}\right\}_{{\bf g}i;{\bf
g}^{\prime}i^{\prime}}E^{out\ -}_{{\bf
g}^{\prime};i^{\prime}}({\bf k}_{\parallel}) \label{ampinm}
\end{equation}
where ${\bf Q}^{{\rm II}}(\nu;1)$ and ${\bf Q}^{{\rm
III}}(\nu-1;2)$ are again the appropriate matrices, shown
schematically in Fig.~\ref{fig1}. A more detailed description of
these matrices and the way these are calculated is to be found in
Ref.~\onlinecite{comphy}. We note that for $\nu=1(N)$ we have only
waves incident from the right (left).

Each plane wave in Eqs.~(\ref{inp}) and (\ref{inm}) can be
expanded in spherical waves about the central sphere of the
$\nu$-th plane in the manner of Eqs.~(\ref{eq:sph_inc}) and
(\ref{eq:j_func}). For a plane wave ${\bf E}^{in\ -}_{{\bf
g}}({\bf k}_{\parallel})\exp[{\rm i}{\bf K}^{-}_{{\bf
g}}\cdot({\bf r}\ -~{\bf A}_{2}(\nu))]$, incident on the $\nu$-th
plane from the right, the multipole coefficients are given by
\cite{comphy}
\begin{equation}
a^{0}_{L}({\bf K}^{-}_{{\bf g}})=\exp [-{\rm i}{\bf K}^{-}_{{\bf
g}}\cdot{\bf d}_{2}(\nu)] \sum_{i}A^{0}_{L;i}({\bf K}^{-}_{{\bf
g}}) E^{in\ -}_{{\bf g};i}({\bf k}_{\parallel}) \label{sphinm}
\end{equation}
And for a plane wave, ${\bf E}^{in\ +}_{{\bf g}}({\bf
k}_{\parallel})\exp[{\rm i}{\bf K}^{+}_{{\bf g}}\cdot({\bf r}-{\bf
A}_{1}(\nu))]$, incident on the $\nu$-th plane from the left, the
multipole coefficients are \cite{comphy}
\begin{equation}
a^{0}_{L}({\bf K}^{+}_{{\bf g}})=\exp [{\rm i}{\bf K}^{+}_{{\bf
g}}\cdot{\bf d}_{1}(\nu)] \sum_{i}A^{0}_{L;i}({\bf K}^{+}_{{\bf
g}}) E^{in\ +}_{{\bf g};i}({\bf k}_{\parallel}). \label{sphinp}
\end{equation}
where $A^{0}_{L;i}$ are given by Eqs.~(12) and (13) of
Ref.~\onlinecite{comphy}.

Finally, to obtain the wave incident on the central sphere of the
$\nu$-th plane, which derives from the outgoing wave of
Eq.~(\ref{sc}), we must add to the waves given by Eqs.~(\ref{inp})
and (\ref{inm}) that which is due to the wave scattered from all
the other spheres of the $\nu$-th plane and it is given by
multiplying the coefficients $a^{0}_{L}$ of Eqs.~(\ref{sphinm})
and (\ref{sphinp}) by the multiple-scattering matrix $\Bigl[[{\bf
I}-{\bf \Omega}{\bf T}_{\nu}]^{-1}\Bigr]_{LL'}$ for the $\nu$-th
plane of spheres. We have
\begin{eqnarray}
\sum_{L'}P^{0n}_{\nu;LL'}b^{+}_{L'}(n;\nu)&=&
\frac{1}{S_{0}}\int\int_{SBZ} d^{2}k_{\parallel} \sum_{{\bf
g}}\sum_{s=\pm}\sum_{L'}\Bigl[[{\bf I}-{\bf \Omega}{\bf
T}_{\nu}]^{-1}\Bigl]_{LL'} a^{0}_{L'}({\bf K}^{s}_{{\bf g}})
\label{inttotal}
\\ &=& \sum_{L'} \frac{1}{S_{0}}\int\int_{SBZ} d^{2}k_{\parallel}
\exp(-{\rm i}{\bf k}_{\parallel}\cdot{\bf R}_{n}) \Bigl[[{\bf
I}-{\bf \Omega}{\bf T}_{\nu}]^{-1}{\bf \Gamma_{\nu}} \Bigr]_{LL'}
b^{+}_{L'}(n;\nu) \nonumber \\  \label{ponn}
\end{eqnarray}
where  $\Gamma_{\nu:LL'}$ is a matrix defined by
\begin{eqnarray}
  \Gamma_{\nu; P l m, P' l^{\prime}m^{\prime}}({\bf
k}_{\parallel};\omega)&=&\sum_{{\bf g},i}\sum_{{\bf
g}^{\prime},i^{\prime}} \Biggl\{ \exp[-{\rm i}({\bf K}^{-}_{{\bf
g}}-{\bf K}^{+}_{{\bf g}^{\prime}})\cdot{\bf d}_{2}(\nu)]A^{0}_{P
l m;i}({\bf K}^{-}_{{\bf g}}) \Bigr. \nonumber
\\ &\times& \Bigl[{\bf Q}^{{\rm III}}(\nu;2)[{\bf I}-{\bf Q}^{{\rm
II}}(\nu+1;1){\bf Q}^{{\rm III}}(\nu;2)]^{-1}\Bigr]_{{\bf g}i;{\bf
g}^{\prime}i^{\prime}}\Delta_{P'l^{\prime}m^{\prime};i^{\prime}}({\bf
K}^{+}_{{\bf g}^{\prime}})   \nonumber
\\ &+&\exp[{\rm i}({\bf K}^{+}_{{\bf
g}}-{\bf K}^{-}_{{\bf g}^{\prime}})\cdot{\bf
d}_{1}(\nu)]A^{0}_{P l m;i}({\bf K}^{+}_{{\bf g}}) \nonumber \\
&\times&\Bigl. \Bigl[{\bf Q}^{{\rm II}}(\nu;1)[{\bf I}-{\bf
Q}^{{\rm III}}(\nu-1;2){\bf Q}^{{\rm II}}(\nu;1)]^{-1}\Bigr]_{{\bf
g}i;{\bf
g}^{\prime}i^{\prime}}\Delta_{P'l^{\prime}m^{\prime};i^{\prime}}({\bf
K}^{-}_{{\bf g}^{\prime}}) \Biggr\} \nonumber \\ \label{glm}
\end{eqnarray}

Therefore, from Eq.~(\ref{ponn}), $P^{0n}_{\nu;LL'}$ is given by
\begin{equation}
P^{0n}_{\nu;LL'}= \frac{1}{S_{0}}\int\int_{SBZ} d^{2}k_{\parallel}
\exp(-{\rm i}{\bf k}_{\parallel}\cdot{\bf R}_{n})\Bigl[[{\bf
I}-{\bf \Omega}{\bf T}_{\nu}]^{-1}{\bf \Gamma_{\nu}} \Bigr]_{LL'}
\label{eq:plm_final}
\end{equation}

Accordingly the second term in Eq.~(\ref{gdef}) becomes
\cite{cpa3d}
\begin{equation}
\sum_{n}\sum_{L''} \sum_{L'''}
P^{0n}_{\nu;LL''}T_{\nu;L''L'''}D^{n0}_{\nu;L'''L'}=\frac{1}{S_{0}}\int\int_{SBZ}
d^{2}k_{\parallel} \Bigl[[{\bf I}-{\bf \Omega}{\bf
T}_{\nu}]^{-1}{\bf \Gamma}_{\nu}{\bf T}_{\nu}{\bf
D}_{\nu}\Bigr]_{LL'} \label{ptd}
\end{equation}
where $D_{\nu;LL'}({\bf k}_{\parallel})$ is given by
Eq.~(\ref{dfourier}). Finally, the matrix $F^{00}_{\nu;LL'}$,
defined by Eq.~(\ref{gdef}), is given by
\begin{equation}
F^{00}_{\nu;LL'}=\frac{1}{S_{0}}\int\int_{SBZ} d^{2}k_{\parallel}
\Bigl[ [{\bf I}-{\bf \Omega}{\bf T}_{\nu}]^{-1} [ {\bf
\Omega}+{\bf \Gamma}_{\nu} ({\bf I}+{\bf T}_{\nu}[{\bf I}-{\bf
\Omega}{\bf T}_{\nu}]^{-1}{\bf \Omega}) ]\Bigr]_{LL'}
\label{gcalc}
\end{equation}

\section{Numerical example}
\label{sec:example}

The evaluation of the propagator, either from Eq.~(\ref{dnm}) or
Eq.~(\ref{gcalc}), requires a numerical integration over the
entire SBZ. Using symmetry to reduce the area of integration to a
part of SBZ is not profitable in the present case. However, when
one deals with scatterers whose dielectric function contains a
positive imaginary part, the intergrand in Eqs.~(\ref{dnm}) or
(\ref{gcalc}) is a relatively smooth function of ${\bf
k}_{\parallel}$, and the integration can be performed without much
difficulty by subdividing the SBZ (a square in our example) into
small squares, within which a nine-point integration formula
\cite{abramo} is very efficient. Using this formula we managed
good convergence with a total of 576 points in the SBZ.

When computing the vdW for $T=0$, we first integrate the Maxwell
stress tensor for a specific frequency over the surface of the
body and afterwards we perform the frequency integration, i.e.,
the vdW force $F$ is calculated by integrating the force spectrum
$F(\omega)$: $F=\int_{0}^{\infty} F(\omega)$. Both integrals are
obtained numerically. We note that, in the Lifshitz theory for
half-spaces, \cite{schwinger} the frequency integration is done
analytically using contour integration. The numerical integral
over frequencies is convergent since, in the limit of $\omega
\rightarrow \infty$, the refractive index of most materials tends
to unity and the corresponding Green's tensor of the system tends
to that of vacuum which is constant in space. However, the
integral over a closed surface of a constant tensor vanishes and
therefore $F(\omega)\rightarrow 0$ as $\omega \rightarrow \infty$.

\begin{figure}[h]
\centerline{\includegraphics[width=12cm]{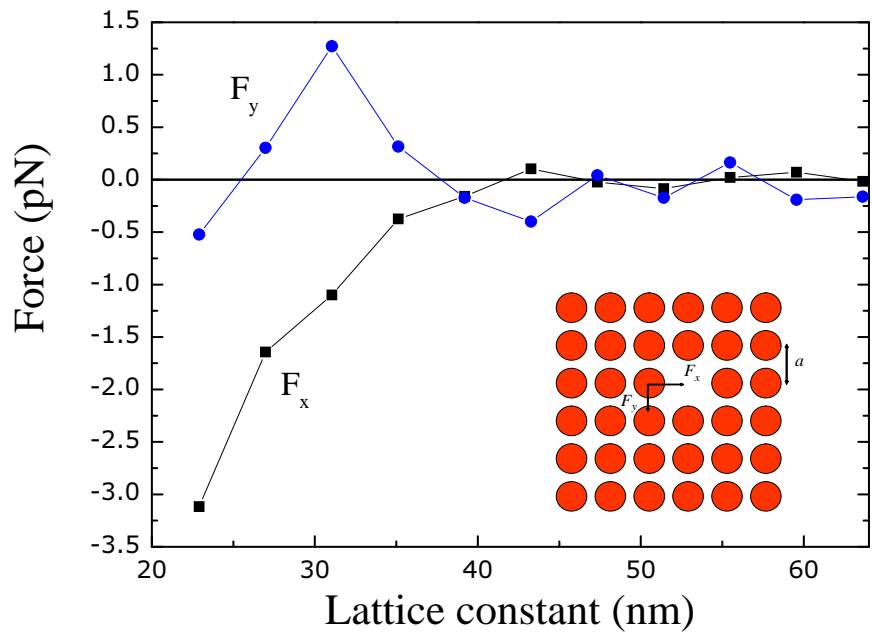}}
\caption{(Color online) Inset: 2D square lattice of 10~nm
polystyrene spheres containing a single defect (one missing
sphere). Graph: the $x$- (squares) and $y$- (circles) component of
the vdW force exerted on a single polystyrene nanosphere when its
right neighboring sphere is missing.} \label{fig2}
\end{figure}

We consider the case of a 2D square lattice (monolayer) of
polystyrene nanospheres of radius 10~nm. The dielectric function
of the spheres which is generally complex for high frequencies is
taken from numerical fit to experimental data.
\cite{parsegian_book} We have calculated the force acting on a
single polystyrene nanosphere when we remove one of its first
neighbouring spheres (see inset of Fig.~\ref{fig2}). In this case,
we first calculate the propagator for the periodic square lattice
from Eqs.~(\ref{dnm}) and (\ref{dfourier}) (which yields vanishing
net vdW force) and then make use of
Eq.~(\ref{eq:d_propagator_defect}).

In Fig.~\ref{fig2} we show the net vdW force ($x$- and $y$-
components) for different lattice constants $a$ of the the
underlying 2D square lattice. While each of the components
oscillates from positive to negative values, it is evident that
there exists a value of the lattice constant $a$, namely $a\simeq
47$~nm, where the net force is zero and this particular sphere
rests in equilibrium. Overall, the magnitude of the vdW force
decreases with the lattice constant, as expected.

\section{Conclusion}
\label{sec:conclusion}

We have presented a method for the calculation of the vdW forces
in colloidal systems such clusters of colloidal particles,
infinite periodic or defected crystals, and colloidal crystals
slabs. The method is based on the fluctuation-dissipation theorem
which relates the cross-spectral correlation functions entering
the formula for the vdW force (integral of the Maxwell stress
tensor over the particle surface) with the EM Green's tensor of
the system of particles (scatterers). The calculation of the
Green's tensor is based on a rigorous multiple-scattering
formalism for EM waves. The accuracy stems from the fact that it
does not include any kind of approximations apart from the
unavoidable cutoffs in the angular momentum expansion and/ or in
the plane-wave expansion of the EM field. As such, the method
includes all essential multipole terms beyond the dipole term in
the EM response of the scatterers and is valid for any distance
between the scatterers. By including {\em a priori} all the
possible multiple-scattering processes of the vacuum fluctuations,
the method, naturally, accounts for all possible many-body
interactions between the scatterers and therefore goes beyond the
approximation of pairwise interactions.

Finally, we note that a theoretical approach, analogous to the
multiple-scattering treatment for the wave equation, has been
developed for solving the Poisson equation in solids described by
arbitrarily shaped, space-filling charges. \cite{gonis} By
combining this electrostatic multiple-scattering approach with the
vdW theory presented in this work, one can devise a general,
first-principles theory for the determination of colloidal
structure.

\small
\normalsize

\small
\normalsize


\end{document}